# Unveiling the distinctive traits of a nation's research performance: the case of Italy and Norway

**Authors:** Giovanni Abramo[1]*, Dag W. Aksnes[2], Ciriaco Andrea D'Angelo[3,1]

**Affiliations:**

[1] Laboratory for Studies in Research Evaluation, Institute for System Analysis and Computer Science (IASI-CNR). National Research Council, Rome, Italy
[2] Nordic Institute for Studies in Innovation, Research and Education, Oslo, Norway
[3] University of Rome "Tor Vergata", Dept of Engineering and Management, Rome, Italy

*corresponding author*

**Abstract**
In this study we are analysing the research performance of Italian and Norwegian professors using constituent components of the Fractional Scientific Strength (FSS) indicator. The main focus is on differences across fields in publication output and citation impact. The overall performance (FSS) of the two countries, which differ considerably in research size and profile, is remarkedly similar. However, an in-depth analysis shows that there are large underlying performance differences. An average Italian professor publishes more papers than a Norwegian, while the citation impact of the research output is higher for the Norwegians. In addition, at field level the pattern varies along both dimensions, and we analyse in which fields each country have their relative strengths. Overall, this study contributes to further insights on how the research performance of different countries may be analysed and compared, to inform research policy.

# 1. Introduction

One of the principal objectives that governments and institutions pursue through research evaluation is improving research performance. Whether associated or not to financial rewards or other competitive mechanisms, the simple communication to individuals of their performance scores and ranks, if properly channelled, can be instrumental to continuous improvement.

All inputs being equal, performance can be increased by producing more research products (alone or in collaboration), or of better quality (higher impact), or both. Providing scientists with information along each single dimension of their overall research performance, might serve better the aim of stimulating improvement. In this way, in fact, the scientist is informed about the dimensions with higher margins of efficiency gains. At large scale level of national scientific systems, governments and policy makers as well might benefit from such information to complement that of comparative world performance rank. It can inform decisions on the dimensions along which to focus and orient policy measures.

Following on our previous study on the comparison of research productivity of Italian and Norwegian academics (Abramo, Aksnes, & D'Angelo, 2020), in this work we present a methodology to measure the single components of research productivity, and apply it to further scan and contrast the two countries' academic systems.

In terms of their science systems, Italy and Norway are rather different. First, the total scientific output of Italy measured by number of journal articles is approximately five times as large as Norway's. In 2018 (latest available data), Italy ranked as the eight largest science producing country in the world while Norway held the 29[th] position (Norges forskningsråd, 2019). However, compared with the population size, the per capita production is 150% higher for Norway than for Italy. Furthermore, the overall citation impact of the research output per capita is also higher for Norway. Thus, Norway may be considered as a more research intensive country than Italy. The underlying reasons for these differences are interesting to analyse further, and the present study aims to contribute to additional knowledge on the issue.

Abramo, Aksnes, and D'Angelo (2020) for the first time ever applied the Fractional Scientific Strength, or FSS, indicator of research performance of professors and universities to a country other than Italy, e.g. Norway. FSS (Abramo & D'Angelo, 2014) differs from the most popular indicators, such as the h-index (Hirsch, 2005), the MNCS (Waltman, Van Eck, Van Leeuwen, Visser, & Van Raan, 2011), and their variants (Wildgaard, Schneider, & Larsen, 2014; Alonso, Cabrerizo, Herrera-Viedmac, & Herrera, 2009; Waltmann, 2016), essentially because it accounts for inputs to research activity. Furthermore, the FSS adopts the publications' fractional counting method, and values them through citation indicators.

We refer the reader to Abramo, Aksnes, and D'Angelo (2020) for details about the academic systems in the two countries, the difficulties to achieve comparable measurements of performance, and the ways to overcome them. There can be found also the procedure to operationalize the measurements, and all the limits and assumptions involved. Where appropriate, we extrapolate from that study, and report in this, the performance scores and ranks in the two countries, across disciplines.

In summary, the previous study showed: i) hardly any differences in the average research productivity of Italian and Norwegian professors; ii) higher concentration of Norwegian professors in the top and the bottom tails of the productivity distribution; and

iii) higher productivity of Norway in Mathematics and Earth and space sciences, and of Italy in Biomedical research and Engineering.

In this study, we complement the above results with more in depth analysis of performance along the single dimensions of overall research productivity of Italian and Norwegian professors. In particular, we contrast the academics of the two countries in terms of yearly output, fractional output, average citations per paper, and average impact factor (IF) per paper.

The elaborations are aimed at answering the following research questions: i) Do the academics of one country tend to publish more than those of the other? ii) In doing so, do they engage in more collaborative work? iii) Do they pursue higher quality research products in terms of citation impact? And, finally iv) Do they publish their results in higher prestigious journals?

In presenting our study, we devote special attention to the definition of each single indicator and the description of the operationalization of their measurement, to make hopeful future replications of the exercise in other countries more straightforward.

From the very beginning, we wish to underline that none of the indicators that we are going to present and measure in this paper can be considered alone performance indicators. Each one represents more or less a dimension of performance, but not the overall performance. They convey complementary information, useful to orient scientists in focusing their efforts for continuous improvement, and managers and policy makers in formulating their interventions to the same aim.

The rest of the paper is structured as follows. In the next section we present the construction of the dataset. In Section 3, we will recall the FSS definition and present those of the complementary indicators we are going to measure. In Section 4, we will present the results of the assessment. Section 5 will conclude the work with the authors' considerations.

2. Data

We observe the research activity of Italian and Norwegian professors in the period 2011-2015.

We extract data on Italian professors from the database on university personnel, maintained by the Italian Ministry of Universities and Research, MIUR. For each professor this database provides information on their name and surname, gender, affiliation, field classification and academic rank, at close of each year.[1]

We extract Norwegian data from a similar database, the Norwegian Research Personnel Register (providing the official Norwegian R&D statistics, compiled by the Nordic Institute for Studies in Innovation, Research and Education - NIFU).

For reasons of significance, the analysis is limited to those professors who held formal faculty positions for at least three years over the 2011-2015 period. Furthermore, the dataset is limited to individuals with at least one publication during the time period (non-publishing personnel is not registered in Norwegian databases).

The dataset used to assess Italian output is extracted from the Italian Observatory of Public Research (ORP), a bibliometric database disambiguated by author developed and maintained by Abramo and D'Angelo, and derived under license from the Clarivate

---
[1] http://cercauniversita.cineca.it/php5/docenti/cerca.php, last accessed on 16 April 2020.

Analytics Web of Science (WoS) Core Collection.

Data on publication output of the Norwegian professors is based on a bibliographic database called Cristin (Current Research Information System in Norway), which is a common documentation system for all institutions in the higher education sector, research institutes and hospitals in Norway.

To pursue distortion-free comparative assessment (see below), we classify each professor in one and only one WoS subject category, SC.[2]

For reasons of significance, we exclude from the analyses professors in SCs belonging to arts and humanities, and few SCs of the social sciences, where the coverage of WoS has largest limitations (Hicks, 1999; Larivière, Archambault, Gingras, Vignola-Gagné, 2006; Aksnes & Sivertsen, 2019).

The various SCs differ considerably in size. In order to avoid possible random fluctuations in performance due to low number of observations, we further exclude those SCs that do not meet the requirement of at least ten professors in total, of both nationalities.

The final dataset consists of 34009 Italian and 4327 Norwegian professors, falling in 177 SCs. Their distribution per academic rank and discipline[3] is shown in Table 1.

*Table 1: Dataset of analysis*

| Discipline | No. of SCs | Italy | | | | Norway | | | |
|---|---|---|---|---|---|---|---|---|---|
| | | Tot. professors | Assistant (%) | Associate (%) | Full (%) | Tot. professors | Assistant (%) | Associate (%) | Full (%) |
| Mathematics | 6 | 2122 | 22.8 | 40.4 | 36.8 | 183 | 2.2 | 29.0 | 68.9 |
| Physics | 16 | 2918 | 23.8 | 43.3 | 32.9 | 256 | 9.8 | 21.5 | 68.8 |
| Chemistry | 7 | 1896 | 28.1 | 43.8 | 28.2 | 122 | 14.8 | 29.5 | 55.7 |
| Earth and Space sciences | 11 | 1873 | 29.6 | 41.6 | 28.8 | 413 | 14.8 | 29.1 | 56.2 |
| Biology | 28 | 5635 | 34.9 | 37.6 | 27.5 | 736 | 20.7 | 28.0 | 51.4 |
| Biomedical research | 14 | 3707 | 37.8 | 36.3 | 25.9 | 245 | 19.2 | 30.2 | 50.6 |
| Clinical medicine | 36 | 7571 | 35.8 | 36.0 | 28.2 | 958 | 10.5 | 31.3 | 58.1 |
| Psychology | 6 | 475 | 31.2 | 39.8 | 29.1 | 148 | 2.7 | 42.6 | 54.7 |
| Engineering | 34 | 5522 | 26.5 | 40.3 | 33.2 | 426 | 5.2 | 27.2 | 67.6 |
| Political and social sciences | 11 | 442 | 20.6 | 41.6 | 37.8 | 438 | 6.2 | 33.1 | 60.7 |
| Economics | 8 | 1848 | 19.0 | 40.4 | 40.6 | 402 | 3.0 | 33.1 | 63.9 |
| *Total* | *177* | *34009* | *30.6* | *39.0* | *30.4* | *4327* | *10.9* | *30.1* | *59.0* |

## 3. The FSS indicator and its components

Productivity is commonly defined as the rate of output per unit of input. It measures how efficiently production inputs are being used. Because publications (output) have different values (impact), and resources employed for research are not homogenous across individuals and organizations, in research systems a more appropriate definition of productivity is: the value of output per Euro spent in research. The FSS indicator is a proxy measure of research productivity. A thorough description of the FSS indicator, and the theory underlying it, can be found in Abramo and D'Angelo (2014).

---

[2] We assigned to each publication the SC or SCs of the hosting journal. We then classified each professor in the most recurrent SC in their publication portfolio. We refer the reader to Abramo, Aksnes, and D'Angelo (2020) for more details on the procedure followed to classify professors.

[3] SCs are grouped in disciplines following a pattern previously published on the website of ISI Journal Citation Reports, but no longer available on the current Clarivate portal. There are no cases in which an SC is assigned to more than one discipline.

To measure the yearly average research productivity of Italian and Norwegian academics, Abramo, Aksnes, and D'Angelo (2020) used the following formula:[4]

$$FSS = \frac{1}{\left(\frac{w_r}{2} + k\right)} \cdot \frac{1}{t} \sum_{i=1}^{N} \frac{c_i}{\bar{c}} f_i \qquad [1]$$

where:
$w_r$ = average yearly salary of professor[5]
$k$ = average yearly capital available for research to professor[6]
$t$ = number of years of work by the professor in period under observation
$N$ = number of publications by the professor in period under observation
$c_i$ = citations received by publication $i$, until 31 October 2018.
$\bar{c}$ = average of distribution of citations received for all WoS cited publications in same year and SC of publication $i$
$f_i$ = fractional contribution of professor to publication $i$.

The fractional contribution equals the inverse of the number of authors in those fields where the practice is to place the authors in simple alphabetical order but assumes different weights in other cases (see Waltman, 2012). For Biology, Biomedical research and Clinical medicine, widespread practice in both Italy and Norway is for the authors to indicate the various contributions to the published research by the order of the names in the bylines. For the above disciplines, we thus give different weights to each co-author according to their position in the list of authors and the character of the co-authorship (intra-mural or extra-mural).[7]

Looking at the formula, it can be seen that productivity is a function of output, or more precisely of individual contribution to output, its quality (impact), and the resources used for production.

In this work we want to assess how Italian professors compare to Norwegian along each dimension of productivity, namely in terms of output, fractional output, average citations per paper, and, although not a dimension of productivity, average IF per paper. In this way we can unveil possible different traits of professors in conducting research activities.

The relevant indicators are the following:

Output (O), average yearly publications authored by the professor, per Euro spent in research:

---

[4] The underlying assumption is that labor and capital equally contribute to production.
[5] We halved labor costs, because we assumed that 50 per cent of professors' time is allocated to activities other than research.
[6] Sources of input data, and assumptions adopted in the measurement can be found in Abramo, Aksnes, and D'Angelo (2020).
[7] If the publication is the outcome of an exclusively intramural collaboration (only one affiliation in the address list), 40% is attributed to both first and last author, and the remaining 20% is divided among all other authors. In contrast, if the publication address list shows extramural collaborations, 30% is attributed to both first and last author; 15% to both second and last but one author; and the remaining 10% is divided among all others. The weighting values were assigned following advice from senior Italian professors in the life sciences. The values could be changed to suit different practices in other national contexts.

$$O = \frac{1}{\left(\frac{w_r}{2} + k\right)} \frac{N}{t}$$

[2]

where:
$w$ = average yearly salary of professor
$k$ = average yearly capital available for research to professor
$N$ = number of publications by the professor in period under observation
$t$ = number of years on staff of professor during the period under observation

Fractional Output (FO), average yearly total contribution to publications authored by the professor, per Euro spent in research:

$$FO = \frac{1}{\left(\frac{w_r}{2} + k\right)} \frac{1}{t} \sum_{i=1}^{N} f_i$$

[3]

where:
$f_i$ = fractional contribution of professor to publication $i$.

Average Citation (AC), average standardized citations per publication:

$$AC = \frac{1}{N} \sum_{i=1}^{N} \frac{c_i}{\bar{c}}$$

[4]

where:
$c_i$ = citations received by publication $i$
$\bar{c}$ = average of distribution of citations received for all WoS cited publications in same year and SC of publication $i$

Average IF (AIF), average standardized IF per publication:

$$AIF = \frac{1}{N} \sum_{i=1}^{N} \frac{IF_i}{\overline{IF}}$$

[5]

where:
$IF$ = IF of journal hosting publication $i$
$\overline{IF}$ = average of distribution of IFs of journals hosting all WoS cited publications in same year and SC of publication $i$

The performance scores of professors belonging to different SCs cannot be compared directly. In fact, i) scientists' intensity of publication remarkably varies across fields, in general (Sandström & Sandström, 2009; Lillquist & Green, 2010; Sorzano, Vargas, Caffarena-Fernández, & Iriarte, 2014), and in both countries in particular (Piro, Aksnes & Rørstad, 2013; D'Angelo & Abramo, 2015); ii) citation behavior varies across fields (Stringer, Sales-Pardo, & Amaral, 2010; Vieira & Gomes, 2010); and iii) the intensity of collaboration, i.e. the average number of co-authors per publication, also varies across fields (Glanzel & Schubert, 2004; Yoshikane & Kageura, 2004; Abramo, D'Angelo, & Murgia, 2013b).

To avoid distortions then, the performance rankings of professors are constructed at SC level.

For comparisons at higher levels of aggregation, i.e. discipline and overall, we normalize performance scores to the average score of all professors of the same SC, but those with nil score.[8] To exemplify, an FSS score of 1.10 means that the professor's performance is 10% above average, in his or her own SC.[9]

In the following tables, figures and text, all performance scores are normalized, but we keep the same denomination for the relevant indicators.

## 4. Results

In the following, we present the comparison of Italian and Norwegian professors, along each of the above indicators, per discipline and overall.

Figure 1 shows the average normalized scores of each indicator, for the overall 34009 Italian and 4327 Norwegian professors. While the two populations show practically the same productivity, FSS,[10] noticeable differences occur for the other indicators. On average, Italian professors publish more (1.4% above average) than Norwegian (11.1% below average). Accounting for the real contribution to each publication, the gap decreases, but it is still there: Italy's FO equals 1.00, while Norway's 0.97.

Because the FSS is the same, it follows that Norway's AC needs to be higher (1.03) than Italy's (1.00). Furthermore, Norwegian academics on average publish in more prestigious journals (AIF = 1.07) than Italian (AIF = 0.99).

In short, while the two countries present the same research productivity, Italians publish more, with larger research teams, while Norwegians publish higher quality products, in more prestigious journals.

---

[8] Abramo, Cicero, and D'Angelo (2012) demonstrated that the average of the distribution (excluding nil values) is the best-performing scaling factor.

[9] Note that the scaling is not referred to world distributions. Since we are comparing Italy vs Norway, the "average" used to rescale original distributions is calculated collapsing Italian and Norwegian performance distributions only.

[10] Note that for FSS, both countries show average figures below one, since in rescaling the FSS of individuals, we exclude nil values. This implies that the average of the overall distribution is below one, in general and for both countries. As for impact indicators (AC, AIF), the effect of removing nil values does not produce the same "visual" effect, while for output indicators (O, FO), we consider only professors with at least one publication authored in the period under observation, consequently there are no nil values.

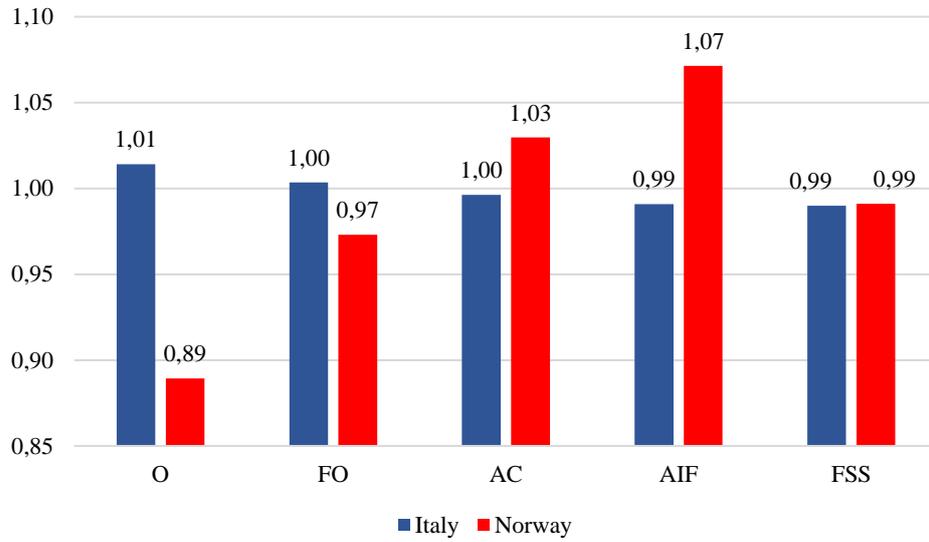

*Figure 1: 2011-2015 average normalized research output and fractional output per Euro spent (O, FO), impact (AC, AIF) and productivity (FSS), at overall country level*

We wondered whether the same traits could be found among top performing scientists as well. Limiting the analysis to top10% professors by FSS at SC level, we observe similar differences for all considered indicators (Figure 2). Norwegian top professors (FSS = 4.15) are slightly more productive than Italian (FSS = 4.10). The first build up their supremacy by producing higher impact publications. In fact, the average O by Italian top professors is 6.8% higher than Norwegian (2.57 vs 2.39), and although it diminishes by FO to 5.2% (2.77 vs 2.62), the roles swap in terms of impact, whereby Norwegian top professors register an AC (AJI) 11.4% (12.4%) higher than Italian.

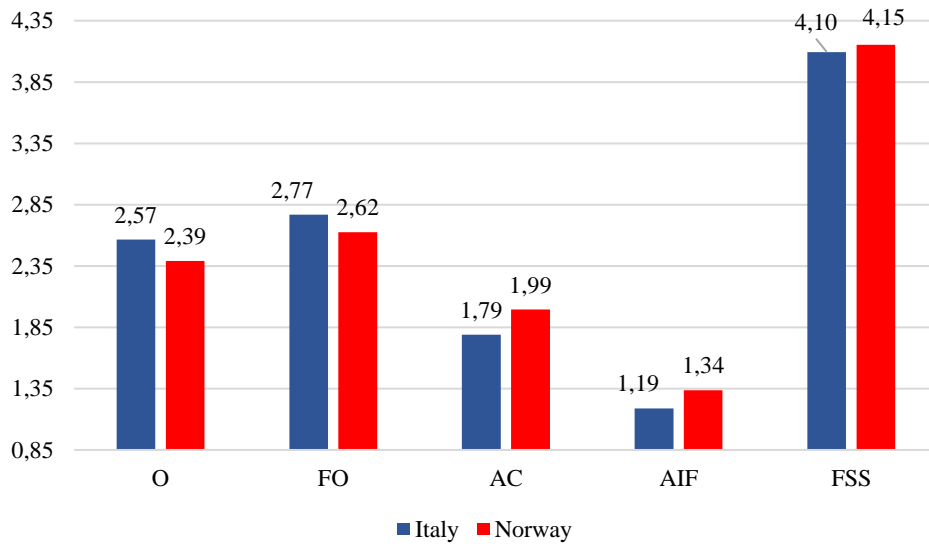

*Figure 2: 2011-2015 average normalized research output and fractional output per Euro spent (O, FO), impact (AC, AIF) and productivity (FSS), at overall country level for top 10% productive professors*

In the following, we present the results at field level (discipline and SC), to assess possible differences across fields between the two populations. We will start with output indicators; impact will follow.

**4.1 Output and fractional output per Euro spent, at field level**

Italian professors' O is higher than Norwegian in all eleven disciplines but Mathematics, as shown in Table 2. The highest gap occurs in Biomedical research, whereby the 3707 Italian professors' O is 1.02, while the 245 Norwegians professors' is 0.77. Noticeable differences occur also in Psychology (1.05 vs 0.84), Biology (1.02 vs 0.82), Engineering (1.01 vs 0.82), and Clinical medicine (1.02 vs 0.85).

When accounting for the real contribution to each paper, Italian professors' contribution (FO) is larger in only five disciplines, especially in Biomedical research (1.01 vs 0.88), and in Biology (1.01 vs 0.90). Norwegians noticeably prevail in Mathematics (1.12 vs 0.99), Chemistry (1.13 vs 0.99), and Earth and space sciences (1.08 vs 0.98).

*Table 2: 2011-2015 normalized average research output and fractional output per Euro spent (O, FO), and productivity (FSS), at discipline level*

| | Italy | | | | Norway | | | |
|---|---|---|---|---|---|---|---|---|
| Discipline | No. of Professors | O | FO | FSS | No. of Professors | O | FO | FSS |
| Mathematics | 2122 | 0.99 | 0.99 | 0.94 | 183 | 1.10 | 1.12 | 1.25 |
| Physics | 2918 | 1.01 | 1.00 | 1.00 | 256 | 0.94 | 1.06 | 0.97 |
| Chemistry | 1896 | 1.01 | 0.99 | 1.00 | 122 | 0.91 | 1.13 | 0.98 |
| Earth and space sciences | 1873 | 1.00 | 0.98 | 0.94 | 413 | 0.98 | 1.08 | 1.27 |
| Biology | 5635 | 1.02 | 1.01 | 1.01 | 736 | 0.82 | 0.90 | 0.90 |
| Biomedical research | 3707 | 1.02 | 1.01 | 1.01 | 245 | 0.77 | 0.88 | 0.89 |
| Clinical medicine | 7571 | 1.02 | 1.01 | 0.99 | 958 | 0.85 | 0.93 | 1.03 |
| Psychology | 475 | 1.05 | 1.02 | 0.98 | 148 | 0.84 | 0.94 | 1.03 |
| Engineering | 5522 | 1.01 | 1.01 | 1.00 | 426 | 0.82 | 0.94 | 0.88 |
| Political and social sciences | 442 | 1.03 | 0.99 | 0.91 | 438 | 0.97 | 1.01 | 0.89 |
| Economics | 1848 | 1.01 | 1.00 | 0.98 | 402 | 0.95 | 1.00 | 0.96 |
| Overall | 34009 | 1.01 | 1.00 | 0.99 | 4327 | 0.89 | 0.97 | 0.99 |

Next, we analyse the research output at the level of SCs. As noted in Section 2, SCs differ considerably in size, and some of them include a rather small number of individuals, particularly for Norway. Nevertheless, an analysis at this level may reveal interesting differences across the nations.

Table 3 shows how the differences by O and FO observed at discipline level vary across the SCs of each single discipline. The FSS scores are reported as a reference. We note that Norwegian professors outperform by FSS Italians in 40% of SCs, in 24% by O, and in 38% by FO. In particular by O, Norwegians never outperform Italians in Psychology, and they do only in 2 of the 28 SCs of Biology, and in 2 of the 14 SCs of Biomedical research. On the other side, in Mathematics, Italians outperform Norwegians in 2 out of 6 SCs. In terms of FO, Norway recovers and, differently from other disciplines, in Chemistry it outperforms Italy in 5 out of 7 SCs.

*Table 3: Number and proportion of fields (SCs) per discipline where 2011-2015 normalized average research output and fractional output per Euro spent (O, FO), and productivity (FSS) of Norway is higher than Italy*

|  | No. of SCs | No. of SCs where Norway outperforms Italy | | |
|---|---|---|---|---|
| Discipline |  | *O* | *FO* | *FSS* |
| Mathematics | 6 | 4 (67%) | 4 (67%) | 3 (50%) |
| Physics | 16 | 7 (44%) | 10 (63%) | 8 (50%) |
| Chemistry | 7 | 2 (29%) | 5 (71%) | 3 (43%) |
| Earth and space sciences | 11 | 4 (36%) | 4 (36%) | 8 (73%) |
| Biology | 28 | 2 (7%) | 7 (25%) | 7 (25%) |
| Biomedical research | 14 | 2 (14%) | 4 (29%) | 5 (36%) |
| Clinical medicine | 36 | 7 (19%) | 11 (31%) | 17 (47%) |
| Psychology | 6 | 0 (0%) | 1 (17%) | 2 (33%) |
| Engineering | 34 | 9 (26%) | 13 (38%) | 11 (32%) |
| Political and social sciences | 11 | 4 (36%) | 4 (36%) | 5 (45%) |
| Economics | 8 | 2 (25%) | 4 (50%) | 2 (25%) |
| Overall | 177 | 43 (24%) | 67 (38%) | 71 (40%) |

Table 4 shows the ten SCs where the gap by O between Norway and Italy is the highest, and vice versa. At the top of the list, in line with the above findings, we find an SC of Mathematics (Mathematics, interdisciplinary applications), where the six Norwegian professors show an O (3.77) about six times as high as that of their 46 Italian counterparts (0.64). The other top SCs fall in four disciplines, namely Engineering (Remote sensing; Construction & building technology), Clinical medicine (Anesthesiology; Substance abuse), Physics (Acoustics; Thermodynamics; Physics, particles & fields), and Political and social sciences (Area studies; Political science). On the other side, the ten SCs with the highest gap in favor of Italy fall in Engineering (7), Clinical medicine (2), and Biomedical research (1).

*Table 4: Top ten fields (SCs) by gap of output (O) in favor of Norway, and the same for Italy*

|  | Italy | | Norway | | |
|---|---|---|---|---|---|
| SC | No. of Prof. | *O* | No. of Prof. | *O* | Δ |
| Mathematics, interdisciplinary applications | 46 | 0.64 | 6 | 3.77 | -3.13 |
| Remote sensing | 102 | 0.92 | 8 | 2.04 | -1.12 |
| Acoustics | 25 | 0.78 | 9 | 1.61 | -0.83 |
| Anesthesiology | 57 | 0.96 | 3 | 1.79 | -0.83 |
| Construction & building technology | 67 | 0.91 | 11 | 1.52 | -0.61 |
| Substance abuse | 9 | 0.69 | 10 | 1.28 | -0.59 |
| Thermodynamics | 97 | 0.98 | 4 | 1.52 | -0.54 |
| Physics, particles & fields | 556 | 0.97 | 30 | 1.50 | -0.52 |
| Area studies | 4 | 0.51 | 73 | 1.03 | -0.52 |
| Political science | 90 | 0.85 | 43 | 1.32 | -0.48 |
| … | | | | | |
| Engineering, multidisciplinary | 11 | 1.13 | 2 | 0.27 | 0.87 |
| Materials science, composites | 59 | 1.03 | 2 | 0.16 | 0.87 |
| Computer science, hardware & architecture | 10 | 1.08 | 1 | 0.19 | 0.89 |
| Computer science, interdisciplinary applications | 44 | 1.17 | 10 | 0.27 | 0.89 |
| Medicine, research & experimental | 66 | 1.03 | 2 | 0.12 | 0.91 |
| Computer science, cybernetics | 6 | 1.39 | 4 | 0.42 | 0.97 |
| Andrology | 9 | 1.10 | 1 | 0.13 | 0.97 |
| Rehabilitation | 47 | 1.47 | 38 | 0.42 | 1.05 |
| Medical informatics | 6 | 1.72 | 6 | 0.28 | 1.43 |
| Information science & library science | 13 | 2.04 | 24 | 0.44 | 1.61 |

Table 5 presents the same analysis by FO. Results are similar to those by O. Six out of top ten SCs, where Norway outperforms Italy, remain the same; while only one changes among those where Italy outperforms

*Table 5: Top ten fields (SCs) by gap of fractional output (FO) in favor of Norway, and the same for Italy*

|  | Italy | | Norway | | |
| --- | --- | --- | --- | --- | --- |
| SC | No. of Prof. | FO | No. of Prof. | FO | Δ |
| Mathematics, interdisciplinary applications | 46 | 0.69 | 6 | 3.41 | -2.72 |
| Hematology | 340 | 0.99 | 4 | 2.19 | -1.21 |
| Remote sensing | 102 | 0.92 | 8 | 2.02 | -1.09 |
| Chemistry, inorganic & nuclear | 242 | 0.98 | 4 | 1.92 | -0.93 |
| Thermodynamics | 97 | 0.96 | 4 | 1.86 | -0.89 |
| Anesthesiology | 57 | 0.96 | 3 | 1.79 | -0.83 |
| Substance abuse | 9 | 0.60 | 10 | 1.36 | -0.75 |
| Instruments & instrumentation | 184 | 0.97 | 7 | 1.71 | -0.74 |
| Reproductive biology | 66 | 0.98 | 2 | 1.58 | -0.60 |
| Acoustics | 25 | 0.85 | 9 | 1.42 | -0.58 |
| … | | | | | |
| Medical informatics | 6 | 1.38 | 6 | 0.62 | 0.76 |
| Computer science, interdisciplinary applications | 44 | 1.14 | 10 | 0.37 | 0.77 |
| Computer science, hardware & architecture | 10 | 1.07 | 1 | 0.27 | 0.81 |
| Medicine, research & experimental | 66 | 1.02 | 2 | 0.20 | 0.82 |
| Engineering, aerospace | 75 | 1.01 | 1 | 0.17 | 0.84 |
| Materials science, composites | 59 | 1.03 | 2 | 0.15 | 0.88 |
| Computer science, cybernetics | 6 | 1.37 | 4 | 0.45 | 0.92 |
| Engineering, multidisciplinary | 11 | 1.15 | 2 | 0.16 | 0.99 |
| Andrology | 9 | 1.11 | 1 | 0.05 | 1.06 |
| Information science & library science | 13 | 1.87 | 24 | 0.53 | 1.35 |

### 4.2 Impact analysis at field level

We now turn to the average impact analysis per discipline. Figure 3 presents the normalized scores of AC. Norwegian academics outperform Italians in six disciplines, most noticeably in Earth and space sciences (0.96 for Italy vs 1.16 for Norway), Mathematics (0.98 vs 1.15) and Clinical medicine (0.99 vs 1.09). Italians outperform Norwegians in five, most noticeably in Engineering (1.01 vs 0.90), Psychology (1.01 vs 0.95), and Physics (1.00 vs 0.95).

Norwegian professors publish on average in more prestigious journals than Italian in eight disciplines, the only exceptions being Physics, Chemistry, and Engineering (Figure 4).

*Figure 3: 2011-2015 normalized average impact of publications (AC), per discipline*

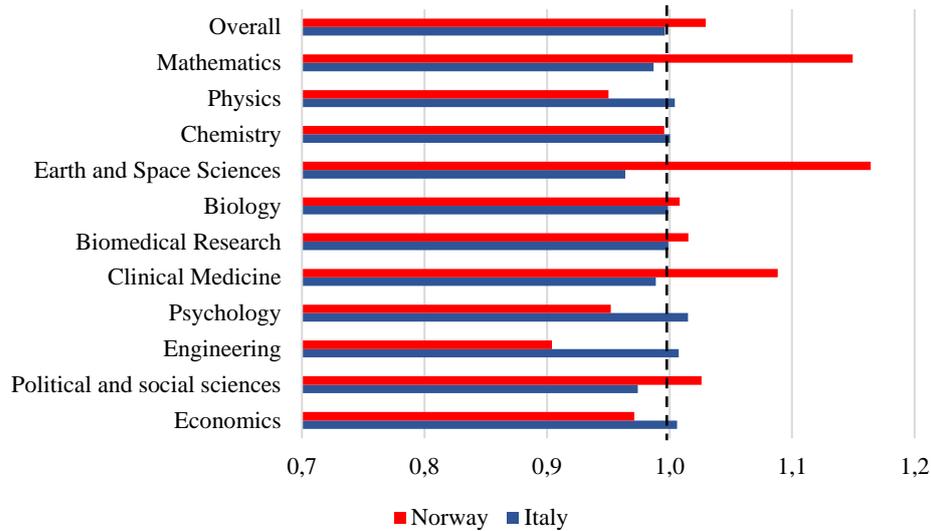

*Figure 4: 2011-2015 normalized average impact factor of journals (AIF), per discipline*

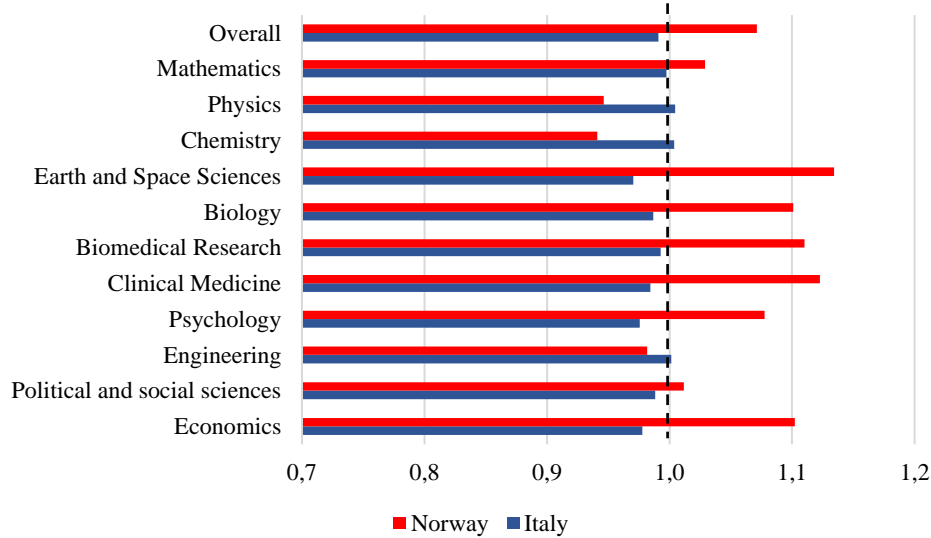

As shown in Figure 1, Norway produces higher impact publications on average, but from Table 6 we see that this occurs in less than 50% of the SCs under observations (86 out of 177). In particular, in 10 of the 11 SCs of Earth and space sciences, in 4 of the 6 SCs of Mathematics, and in 7 of the 11 SCs of Political and social sciences. Viceversa Italians outperform Norwegians in 6 of the 8 SCs of Economics, 5 of the 7 SCs of Chemistry (5 out of 7), and in 10 of the 14 SCs of Biomedical research.

Norwegian research products are published in higher IF journals in 112 SCs (63%).

*Table 6: : Number of fields (SCs) and proportion per discipline where 2011-2015 average citations per publication (AC), average IF per publication (AIF), and productivity (FSS) of Norway is higher than Italy*

| Discipline | No. of SCs | No. of SCs where Norway outperforms Italy | | |
|---|---|---|---|---|
| | | AC | AIF | FSS |
| Mathematics | 6 | 4 (67%) | 5 (83%) | 3 (50%) |
| Physics | 16 | 6 (38%) | 6 (38%) | 8 (50%) |
| Chemistry | 7 | 2 (28%) | 3 (43%) | 3 (43%) |
| Earth and space sciences | 11 | 10 (91%) | 10 (91%) | 8 (73%) |
| Biology | 28 | 16 (57%) | 21 (75%) | 7 (25%) |
| Biomedical research | 14 | 4 (29%) | 11 (79%) | 5 (36%) |
| Clinical medicine | 36 | 20 (56%) | 28 (78%) | 17 (47%) |
| Psychology | 6 | 3 (50%) | 4 (67%) | 2 (33%) |
| Engineering | 34 | 12 (35%) | 15 (44%) | 11 (32%) |
| Political and social sciences | 11 | 7 (64%) | 5 (45%) | 5 (45%) |
| Economics | 8 | 2 (25%) | 4 (50%) | 2 (25%) |
| Overall | 177 | 86 (49%) | 112 (63%) | 71 (40%) |

Table 7 reports the ten SCs with the highest difference of AC for each country. In five SCs the average AC of Norwegian professors is above 2, while for Italians is below 1. The highest gap is registered in History & philosophy of science, with an average AC for 62 Italian professors of 0.74, against 2.59 for 10 Norwegians. Out the top ten SCs where Italy outperforms Norway, five SCs fall in Engineering (Robotics; Engineering, aerospace; Transportation science & technology; Materials science, composites; Computer science, Cybernetics), and three in Economics/Political and social sciences (Public administration; Urban studies; Area studies).

*Table 7: Top ten fields (SCs) by gap of average citation per publication (AC) in favor of Norway, and the same for Italy*

| | Italy | | Norway | | |
|---|---|---|---|---|---|
| SC | No. of Prof. | AC | No. of Prof. | AC | Δ |
| History & philosophy of science | 62 | 0.74 | 10 | 2.59 | -1.84 |
| Engineering, multidisciplinary | 11 | 0.79 | 2 | 2.13 | -1.33 |
| Urology & nephrology | 234 | 0.98 | 3 | 2.21 | -1.23 |
| Medicine, legal | 92 | 0.94 | 5 | 2.10 | -1.16 |
| Anesthesiology | 57 | 0.94 | 3 | 2.06 | -1.12 |
| Statistics & probability | 390 | 0.94 | 29 | 1.83 | -0.89 |
| Remote sensing | 102 | 0.95 | 8 | 1.66 | -0.71 |
| Mathematics, interdisciplinary applications | 46 | 0.92 | 6 | 1.62 | -0.70 |
| Medicine, general & internal | 18 | 0.97 | 1 | 1.62 | -0.65 |
| Physics, atomic, molecular & chemical | 98 | 0.94 | 12 | 1.52 | -0.58 |
| … | | | | | |
| Robotics | 56 | 1.01 | 1 | 0.43 | 0.58 |
| Engineering, aerospace | 75 | 1.01 | 1 | 0.39 | 0.61 |
| Public administration | 21 | 1.26 | 12 | 0.55 | 0.71 |
| Urban studies | 26 | 1.14 | 6 | 0.37 | 0.77 |
| Transportation science & technology | 41 | 1.06 | 3 | 0.24 | 0.81 |
| Physics, multidisciplinary | 157 | 1.05 | 11 | 0.23 | 0.82 |
| Materials science, composites | 59 | 1.03 | 2 | 0.18 | 0.85 |
| Computer science, cybernetics | 6 | 1.35 | 4 | 0.48 | 0.86 |
| Developmental biology | 22 | 1.04 | 1 | 0.13 | 0.91 |
| Area studies | 4 | 1.98 | 73 | 0.95 | 1.04 |

Table 8 proposes the same view by the average IF of hosting journals per paper. Along this indicator, five of the top ten SCs where Norway outperforms Italy fall in Clinical medicine (Andrology; Urology & nephrology; Medicine, general & internal; Anesthesiology; Medicine, legal). Simmetrically, five of the top SCs for Italy fall in Engineering (Robotics; Medical informatics; Computer science, interdisciplinary applications; Nanoscience & nanotechnology; Transportation science & technology).

*Table 8: Top ten fields (SCs) by gap of average journal impact (AIF) in favor of Norway, and the same for Italy*

|  | Italy | | Norway | | |
|---|---|---|---|---|---|
| SC | No. of prof. | *AIF* | No. of prof. | *AIF* | Δ |
| History & philosophy of science | 62 | 0.82 | 10 | 2.14 | -1.32 |
| Andrology | 9 | 0.92 | 1 | 1.73 | -0.82 |
| Urology & nephrology | 234 | 0.99 | 3 | 1.72 | -0.73 |
| Agriculture, multidisciplinary | 67 | 0.96 | 4 | 1.61 | -0.65 |
| Business, finance | 101 | 0.90 | 20 | 1.53 | -0.63 |
| Medicine, general & internal | 18 | 0.97 | 1 | 1.54 | -0.58 |
| Anesthesiology | 57 | 0.97 | 3 | 1.54 | -0.57 |
| Metallurgy & metallurgical engineering | 47 | 0.92 | 9 | 1.44 | -0.52 |
| Medicine, legal | 92 | 0.97 | 5 | 1.49 | -0.52 |
| Horticulture | 162 | 0.99 | 2 | 1.50 | -0.50 |
| … | | | | | |
| Robotics | 56 | 1.01 | 1 | 0.72 | 0.29 |
| Medical informatics | 6 | 1.15 | 6 | 0.85 | 0.29 |
| Communication | 25 | 1.18 | 34 | 0.87 | 0.32 |
| Biophysics | 17 | 1.02 | 1 | 0.69 | 0.33 |
| Computer science, interdisciplinary applications | 44 | 1.06 | 10 | 0.72 | 0.34 |
| Nanoscience & nanotechnology | 32 | 1.01 | 1 | 0.65 | 0.36 |
| Medicine, research & experimental | 66 | 1.01 | 2 | 0.62 | 0.40 |
| Transportation science & technology | 41 | 1.03 | 3 | 0.60 | 0.43 |
| Physics, multidisciplinary | 157 | 1.03 | 11 | 0.53 | 0.50 |
| Developmental biology | 22 | 1.02 | 1 | 0.46 | 0.56 |

## 6. Discussion and conclusions

Cross national comparisons of research performance is a main area of application of bibliometrics. For example, such analyses have a central position in science and technology indicator reports for monitoring scientific development and performance (see e.g. National Science Board, 2020; OECD, 2019; Norges forskningsråd, 2019). Through the large attention these reports receive, the bibliometric indicators play an important role in the public perception of the scientific performance of countries. Our study adds new insight on more advanced methods for comparing national research performance than what is provided through such reports.

Despite Italy and Norway differ considerably along such dimensions as research size and profile, and organisation of the research systems, there are hardly any differences at all in the average research productivity (FSS) of Italian and Norwegian professors (Abramo, Aksnes, & D'Angelo, 2020).

In this work, we delve into the various dimensions of research performance. Overall, the study shows that an average Italian professor publishes more papers than a Norwegian. This difference is quite large when measuring publications on whole count basis but is smaller using fractionalised measures (i.e. in which credit for a publication is allocated according to the contribution of the participating co-authors). We infer that Italian publications must have more co-authors, which suggests that Italians tend to work in larger teams. This might be related to the fact that the Italian research system is much larger than the Norwegian. Possibly, Italian researchers may benefit from larger size universities which can favour larger institutional collaborations. The different collaboration behaviour of Italian and Norwegian professors might reveal an interesting topic of future investigation.

The discussion on the appropriateness of whole versus fractionalised measures of publication output has a long history in bibliometrics, dating back the dispute on the decline of British science during the 1980s (Braun, Glänzel, & Schubert, 1989; Irvine, Martin, Peacock, & Turner, 1985). Moed (2005) argues that the two measurement principles should be seen as complementary where whole counts measure the participation and fractionalised counts the number of creditable papers. Our analysis shows that the different measurement principles yield quite different results, which has also been documented in previous studies (Gauffriau, Olesen Larsen, Maye, Roulin-Perriard, & von Ins, 2008; Aksnes, Sivertsen, Van Leeuwen, & Wendt, 2016). Currently, fractionalised measurement seems to have gained increased popularity (Waltman & Van Eck, 2015), and this is also the principle underlying the FSS indicator.

While the number of publications produced per individual is higher on average for Italian than for Norwegian professors, the pattern is opposite when looking at the citation impact and journal profile. We observe national patterns where Italian academics produce more but with lower impact and in less prestigious journals. Thus, there is a quantity versus quality tension in the national publication patterns, where the first dimension apparently is more emphasised by the Italians and the latter by the Norwegians. However in the combined FSS-indicators these differences tend to level out.

We are not able to provide any final answer on possible reasons underlying these differences. The quality of the research is one obvious dimension, but there are also mechanisms in the research systems that might favour certain publication behaviours more than others. In Norway, there is a performance-based funding system where a bibliometric indicator is applied for the allocation of funding across institutions (Sivertsen, 2018). Although this model is designed to work at an overall national level, previous research has shown that it is sometimes applied at lower levels and may have an incentive effect at the level of individual researchers (Aagaard, 2015). Two important elements of the model should be emphasised. First, extra points are given for publications in the most prestigious publication channels; second, all journal and book publications in accredited publication channels are included, not only publications indexed in WoS . This means that there is an incentive for publishing in prestigious publication outlets and no incentive for publishing in WoS journals specifically.

In Italy, a proportion of public funds started being allocated to universities on the basis of the outcome of the second national research assessment exercise, VQR 2006-2010.[11] The evaluation was based on a restricted number of research products submitted by universities, and the bibliometric indicator applied to assess their relative quality was a combination of citations and IF. The underlying incentive seems to have professors focus on the production of few very high quality outputs, to be published in prestigious journals. In the same period, the "national scientific habilitation" (ASN) for university appointments was introduced in Italy.[12] The ASN required passing the threshold values in all or part of (depending on the field) three bibliometric indicators, namely number of articles, number of citations, contemporary h-index (Sidiropoulos, Katsaros, & Manolopoulos, 2007).

---

[11] https://www.anvur.it/en/homepage/, last accessed on 16 April 2020.
[12] https://abilitazione.miur.it/public/index.php?lang=eng, last accessed on 16 April 2020.

Possibly, Norwegian stronger incentives for quality of research can partly explain why we find the differences across the two nations.

Furthermore, previous research has shown that there is a language bias in the coverage of Web of Science which affects various countries differently (Van Leeuwen, Moed, Tijssen, Visser, & Van Raan, 2001). Although journals in national languages as a general rule are not covered by Web of Science, at least not in the core collection, there are many exceptions. While coverage of national journals will increase the publication numbers of a country, it will have an opposite effect on the citation impact as articles in these journals generally tend to be little cited. There are hardly any national Norwegian journals indexed in the core collection of Web of Science, while for Italy there are several and more and more numerous, especially after the introduction of the VQR. Possibly, this factor too might explain some of the observed country differences across the indicators analysed.

In interpreting the results of the performance analysis, all the usual limits, caveats, assumptions and qualifications of evaluative scientometrics apply, in particular: i) publications are not representative of all knowledge produced; ii) bibliometric repertories applied (WoS) do not cover all publications; and iii) citations are not always certification of real use and representative of all use. Furthermore, results are sensitive to the classification schemes adopted for both publications and professors. Finally, the limited availability of comparable input data required the adoption of few assumptions. Subject to the availability of input data, the present study can be replicated to include other countries.

Studies of the output dimension in other countries would be of particular interest. For a long time various normalization procedures have been applied for calculating citation indicators (Moed, 2005). However, there are no cross-national reference standards for comparing publication output, although previous studies have shown the field dimension to be of particular importance. For example, a study of publication patterns in social sciences and humanities in eight European countries, showed large differences across fields but also across nations (Kulczycki et al., 2018). In addition, publication output has been shown to be influenced by individual variables such as gender (Abramo, D'Angelo, & Caprasecca, 2009; Sugimoto, Lariviere, Ni, Gingras, & Cronin, 2013; Elsevier, 2020), age (Levin & Stephan, 1989; Kyvik, 1990; Gingras, Larivière, Macaluso, & Robitaille, 2008; Abramo, D'Angelo & Murgia), and academic rank (Abramo, D'Angelo, & Di Costa, 2011; Blackburn, Behymer, & Hall, 1978; Ventura & Mombrù, 2006; ). Thus, further research would be required in order to provide better fundaments for cross-national analyses along this dimension of performance.

Comparative research assessment data provides governments, universities, industry and prospective students with valuable information about research performance in a country's higher education institutions. In particular, the in-depth investigation along the single dimensions of research performance, can reveal those with higher potential efficiency gains, and therefore orient the incentive systems for continuous improvement.

The results of the current comparison offer a number of stimuli to further delve into the performance analyses. In addition to the above said collaboration behavior, gender issues might be explored as well. In fact, the two countries present a quite different history of female emancipation, and attitudes concerning the role of women in the family, in the workplace, and in the society in general. It would be interesting then to explore whether gender representation across academic ranks, and differences in performance occur to the same extent in the two countries.